\documentclass[12pt]{article}
\usepackage{epsfig}
\usepackage{axodraw}
\newcommand{\mysection}{\setcounter{equation}{0}\section}

\def\beq{\begin{equation}}
\def\eeq{\end{equation}}
\def\beqa{\begin{eqnarray}}
\def\eeqa{\end{eqnarray}}

\begin{document}

\begin{center}
{\Large \bf Next-to-next-to-leading-order collinear and soft gluon corrections for $t$-channel single top quark production}
\end{center}
\vspace{2mm}
\begin{center}
{\large Nikolaos Kidonakis}\\
\vspace{2mm}
{\it Kennesaw State University, Physics \#1202,\\
1000 Chastain Rd., Kennesaw, GA 30144-5591}\\
\end{center}

\begin{abstract}
I present the resummation of collinear and soft gluon corrections to single  
top quark production in the $t$ channel at next-to-next-to-leading 
logarithm (NNLL) accuracy using two-loop soft anomalous dimensions.
The expansion of the resummed cross section yields approximate 
next-to-next-to-leading order (NNLO) cross sections. 
Numerical results for $t$-channel single top quark 
(or single antitop) production at the Tevatron and the LHC 
are presented, including the dependence of the cross sections on 
the top quark mass, and the uncertainties from scale variation and 
parton distributions. Combined results for all single top quark production 
channels are also given.

\end{abstract}

\mysection{Introduction}

The observation of single top quark production at the Tevatron 
\cite{D0st,CDFst,Tevew,AHTJ} and the re-discovery of top quarks at the LHC 
\cite{CMS, ATLAS} has 
increased the need for theoretical calculations of the cross sections for 
the relevant processes. 
The single top cross section is less than half of that for $t{\bar t}$ production 
while the backgrounds are considerable and make the extraction of the signal challenging.  
Single top quark production is important in probing electroweak theory and discovering new
physics since the top quark mass is of the same order of magnitude 
as the electroweak symmetry breaking scale, and it
provides opportunities for the study of the 
electroweak properties of the top quark.

Single top quarks can be produced through three distinct 
partonic processes. One of them is the $t$-channel process that proceeds via
the exchange of a space-like $W$ boson (Fig. 1), a second is the $s$-channel 
process that proceeds via the exchange of a time-like $W$ boson, 
and a third is associated $tW$ production (and the related $tH^-$ production). 
At both the LHC and the Tevatron the $t$-channel is numerically dominant.
The $t$-channel partonic processes are of the form $qb \rightarrow q' t$ 
and ${\bar q} b \rightarrow {\bar q}' t$.

\begin{figure}
\begin{center}
\includegraphics[width=8cm]{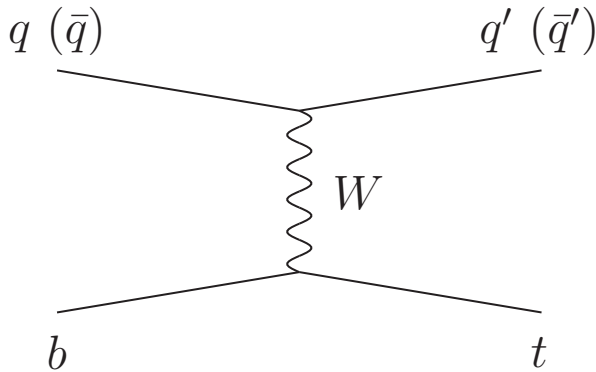}
\caption{\label{tlo} Leading-order $t$-channel diagram for 
single top quark production.}
\end{center}
\end{figure}

Calculations of next-to-leading order (NLO) corrections for $t$-channel production at the 
differential level have been known for some time \cite{BWHL} and recent 
updates and further studies of the NLO cross section have appeared in 
\cite{CFMT,FMS,SYMC,FGMS}. 
Theoretical calculations for single top quark production 
beyond NLO that include higher-order corrections 
from next-to-leading-logarithm (NLL) soft-gluon resummation 
appeared in \cite{NKcH,NKsingletopTev,NKsingletopLHC} for the three 
channels.  

Recent advances in two-loop calculations with both massless \cite{ADS} and massive \cite{NK2l}
quarks now allow next-to-next-to-leading-logarithm (NNLL) resummation.
More recently, calculations at NNLL accuracy have appeared for $s$-channel production 
\cite{NKsch} and $tW$ or $tH^-$ production \cite{NKtWH}. Related NNLL calculations for 
$t{\bar t}$ production appeared in \cite{NKttb}. The present paper completes the NNLL results 
for top quark processes by studying $t$-channel single top quark production.

A different formalism for resummation based on soft-collinear effective 
theory (SCET) was used for $s$-channel production in Ref. \cite{ZLWZ} 
and for $t$-channel production in Ref. \cite{WLZZ}.
Different choices for the threshold kinematics variables were used in 
\cite{ZLWZ,WLZZ} than in \cite{NKsch} and here. 
The authors of \cite{ZLWZ} have argued that their kinematics choice is 
more physical. However, they find that the soft-gluon contributions 
in the $s$ channel are small 
and their threshold expansion at NLO is not a good approximation to the 
exact NLO cross section at LHC energies. They also find smaller 
higher-order enhancements 
from soft gluons than in \cite{NKsch}. We disagree with their choices and conclusions (also, contrary to what is stated in \cite{ZLWZ}, our expressions 
in \cite{NKsch} included the complete NNLL terms).
 
In our approach the soft-gluon contributions are dominant, in both $s$ and $t$
channels, and hence the resummation is more relevant and the NLO threshold 
expansion is a good approximation to the exact NLO result at both 
Tevatron and LHC energies. Hence, our complete NNLL higher-order corrections are also 
expected to be a better approximation. 
In this paper we use the same general formalism as presented in 
\cite{NKsingletopTev,NKsch}. Our new $t$-channel results at NNLL accuracy 
improve and update the earlier results in \cite{NKsingletopTev,NKsingletopLHC}.

In Section 2 we briefly describe the threshold resummation formalism and 
provide expressions for the two-loop soft anomalous dimension that leads to 
a NNLL resummed cross section. We 
expand the resummed cross section in powers of $\alpha_s$ and provide formulas 
for the soft-gluon corrections through next-to-next-to-leading order (NNLO). 
In Section 3 we present numerical results for single top quark 
(or single antitop) production via the $t$ channel at the Tevatron. 
Analogous results are provided for single top production at the LHC in Section 4 and 
for single antitop production at the LHC in Section 5. We conclude in Section 6 
with a combination of the results in different channels for both Tevatron and 
LHC energies.

\mysection{Two-loop resummation}

In this section we present the resummed cross section for 
$t$-channel single top quark production.  
Details of the general resummation formalism for this process 
have been presented in \cite{NKsingletopTev}.

For the partonic process $q(p_1)+b(p_2) \rightarrow q'(p_3)+t(p_4)$,
the kinematical invariants are 
$s=(p_1+p_2)^2$, $t=(p_1-p_3)^2$, $u=(p_2-p_3)^2$, 
$s_4=s+t+u-m_t^2$, where $m_t$ is the top quark mass and we ignore the mass 
of the $b$-quark.
Near the threshold of the partonic energy to produce the final state 
with the top quark, the quantity $s_4$, which measures distance from threshold, 
goes to zero.
The threshold corrections then take the form of logarithmic plus 
distributions, $[\ln^l(s_4/m_t^2)/s_4]_+$, 
where $l\le 2n-1$ for the $n$-th order QCD corrections.

The resummation of the threshold logarithms is a consequence 
of the factorization of the cross section into hard, soft, 
and jet functions that describe, respectively, the hard scattering, 
noncollinear soft gluon emission, and collinear gluon emission from the 
partons in the process \cite{KS}.
We take moments of the partonic cross section,  
${\hat\sigma}(N)=\int (ds_4/s) \;  e^{-N s_4/s} {\hat\sigma}(s_4)$.
The moments of the logarithms of $s_4$ yield 
logarithms of the moment variable $N$, which exponentiate.
The resummed partonic cross section in moment space 
is then   
\beqa
{\hat{\sigma}}^{res}(N) &=&   
\exp\left[ \sum_{i=1,2} E(N_i)\right] \; 
\exp\left[ {E'}(N')\right] \; 
\exp \left[\sum_{i=1,2} 2 \int_{\mu_F}^{\sqrt{s}} \frac{d\mu}{\mu}\;
\gamma_{q/q}\left({\tilde N}_i, \alpha_s(\mu)\right)\right] \;
\nonumber\\ && \hspace{-10mm} \times \,
{\rm Tr} \left \{H\left(\alpha_s(\sqrt{s})\right) \;
\exp \left[\int_{\sqrt{s}}^{{\sqrt{s}}/{\tilde N'}} 
\frac{d\mu}{\mu} \;\Gamma_S^{\dagger}
\left(\alpha_s(\mu)\right)\right] \right.
\nonumber\\ && \quad \left.  \times \,
S\left(\alpha_s(\sqrt{s}/{\tilde N'})
\right) \; \exp \left[\int_{\sqrt{s}}^{{\sqrt{s}}/{\tilde N'}} 
\frac{d\mu}{\mu}\; \Gamma_S
\left(\alpha_s(\mu)\right)\right] \right\} \, .
\label{resHS}
\eeqa

In Eq. (\ref{resHS}) the first exponent resums collinear and soft gluon emission \cite{GS87,CT89}
from the initial-state partons and it 
is given in the $\overline{\rm MS}$ scheme by
\beq
E(N_i)=
\int^1_0 dz \frac{z^{N_i-1}-1}{1-z}\;
\left \{\int_1^{(1-z)^2} \frac{d\lambda}{\lambda}
A\left(\alpha_s(\lambda s)\right)
+D\left[\alpha_s((1-z)^2 s)\right]\right\} \, .
\label{Eexp}
\eeq
Here $N_1=N[(m_t^2-u)/m_t^2]$ and 
$N_2=N[(m_t^2-t)/m_t^2]$. 
The quantity $A$ has a perturbative expansion, 
$A=\sum_n (\alpha_s/\pi)^n A^{(n)}$. 
Here 
$A^{(1)}=C_F$ with $C_F=(N_c^2-1)/(2N_c)$ where $N_c=3$ is 
the number of colors,
while $A^{(2)}=C_F K/2$ with 
$K= C_A\; ( 67/18-\pi^2/6 ) - 5n_f/9$ \cite{KT},
where $C_A=N_c$, and $n_f=5$ is the number of light quark flavors. 

Also $D=\sum_n (\alpha_s/\pi)^n D^{(n)}$, 
where in Feynman gauge $D^{(1)}=0$ and \cite{CLS}
\beq
D^{(2)}=C_F C_A \left(-\frac{101}{54}+\frac{11}{6} \zeta_2
+\frac{7}{4}\zeta_3\right)
+C_F n_f \left(\frac{7}{27}-\frac{\zeta_2}{3}\right) 
\eeq
where $\zeta_2=\pi^2/6$ and $\zeta_3=1.2020569\cdots$.

The second exponent in Eq. (\ref{resHS}) resums soft and collinear corrections 
\cite{GS87,CT89,LOS,NKVDD}
from the final-state massless quark and it is given by
\beq
{E'}(N')=
\int^1_0 dz \frac{z^{N'-1}-1}{1-z}\;
\left \{\int^{1-z}_{(1-z)^2} \frac{d\lambda}{\lambda}
A \left(\alpha_s\left(\lambda s\right)\right)
+B\left[\alpha_s((1-z)s)\right]
+D\left[\alpha_s((1-z)^2 s)\right]\right\} \, ,
\label{E'exp}
\eeq
where  $N'=N (s/m_t^2)$ and $A$ and $D$ are defined above. 
Here $B=\sum_n (\alpha_s/\pi)^n B^{(n)}$ 
with $B^{(1)}=-3C_F/4$ and
\beq
B^{(2)}=C_F^2\left(-\frac{3}{32}+\frac{3}{4}\zeta_2-\frac{3}{2}\zeta_3\right)
+C_F C_A \left(-\frac{1539}{864}-\frac{11}{12}\zeta_2+\frac{3}{4}\zeta_3\right)
+n_f C_F \left(\frac{135}{432}+\frac{\zeta_2}{6}\right).
\eeq

In the third exponent the parton-density anomalous dimension $\gamma_{q/q}$ 
controls the factorization scale, $\mu_F$, dependence of the cross section. 
We write $\gamma_{q/q}=-A \ln {\tilde N}_i +\gamma_q$ where $A$ was 
defined above, ${\tilde N}_i=N_i e^{\gamma_E}$ 
with $\gamma_E$ the Euler constant, and 
$\gamma_q=\sum_n (\alpha_s/\pi)^n \gamma_q^{(n)}$
where  $\gamma_q^{(1)}=3C_F/4$.

$H$ is the hard-scattering function 
while $S$ is the soft function describing noncollinear soft gluon emission 
\cite{KS}.
The evolution of the soft function is controlled by the soft anomalous 
dimension $\Gamma_S$.
The functions $H$, $S$, and $\Gamma_S$  are matrices in a basis 
consisting of color exchange, and we take the trace of the product involving 
these matrices in Eq. (\ref{resHS}).
 For the $t$-channel process with color indices 
$a+b \rightarrow c+d$ we choose the color basis 
$e_1=\delta_{ac}\delta_{bd}$ and 
$e_2=T^e_{ca} T^e_{db}$.
We write the perturbative series for the soft anomalous dimension as
$\Gamma_S=\sum_n (\alpha_s/\pi)^n \Gamma_S^{(n)}$. 
Because of the simple color structure of the hard scattering
for single top $t$-channel production, 
the hard and soft matrices take a very simple form and only the first 
diagonal element of the one-loop soft anomalous dimension matrix, 
$\Gamma_{S\, 11}^{(1)}$, is needed in the NNLO expansion at NLL accuracy. 

By expanding the resummed cross section, Eq. (\ref{resHS}), 
in powers of $\alpha_s$ we derive fixed-order corrections, thus avoiding  
the prescription ambiguity that a fully resummed cross section entails 
to avoid the infrared singularity. This has been our approach 
in Refs. \cite{NKcH,NKsingletopTev,NKsingletopLHC,NKsch,NKtWH,NKttb}.

The NLO soft-gluon corrections to the differential cross section are
\beq
\frac{d^2{\hat\sigma}^{(1)}}{dt \, du}
=F^B
\frac{\alpha_s(\mu_R)}{\pi} \left\{
c_{3} \left[\frac{\ln(s_4/m_t^2)}{s_4}\right]_+
+c_{2} \left[\frac{1}{s_4}\right]_+ \right\} \, ,
\label{NLO}
\eeq
where $F_B$ is the Born term \cite{NKsingletopTev} 
and $\mu_R$ is the renormalization scale.
The leading coefficient is $c_{3}=3C_F$.  
The next-to-leading coefficient, $c_2$, can be written as 
$c_2=T_2+c_2^{\mu}$, 
where $T_2$ represents the scale-independent part of $c_2$ and  
$c_2^{\mu}$ has all the scale dependence.
Here
\beq
T_{2}=2 {\Gamma}_{S\, 11}^{(1)}
-\frac{3}{4}C_F-2C_F\ln\left(\frac{(t-m^2_t)(u-m^2_t)}{m_t^4}\right)
-3C_F\ln\left(\frac{m_t^2}{s}\right)
\eeq
and
\beq
c_2^{\mu}= -2C_F \ln\left(\frac{\mu_F^2}{m_t^2}\right) \, .
\eeq

The required element $\Gamma_{S\, 11}^{(1)}$ of the one-loop soft 
anomalous dimension matrix for $t$-channel single-top production, necessary 
for NLL accuracy, was calculated 
first in Ref. \cite{NKsingletopTev} in axial gauge. 
The calculation involves one-loop eikonal diagrams with vertex corrections and 
a self-energy correction for the top quark line.
The soft anomalous dimension is determined 
from the coefficients of the ultraviolet poles in dimensional regularization.
In this paper we use the Feynman gauge and thus the result takes the slightly 
different form
\beq
{\Gamma}_{S\, 11}^{(1)}
=C_F \left[\ln\left(\frac{-t}{s}\right)
+\ln\left(\frac{m_t^2-t}{m_t\sqrt{s}}\right)-\frac{1}{2}\right]\, . 
\label{Gamma111}
\eeq
This change is of course compensated by different expressions for the 
$D$ term in Eqs. (\ref{Eexp}) and (\ref{E'exp}) than those used in 
\cite{NKsingletopTev} so that the final result for the resummed 
cross section is identical in the two gauges, as expected.

If we denote by $c_1$ the coefficient of the $\delta(s_4)$ corrections
in the NLO cross section, 
then we can calculate the scale-dependent part, $c_1^{\mu}$, from our 
resummation formalism
\beq
c_1^{\mu}=\left[C_F \ln\left(\frac{(t-m^2_t)(u-m^2_t)}{m_t^4}\right)
-\frac{3}{2}C_F\right]\ln\left(\frac{\mu_F^2}{m_t^2}\right) \, .
\eeq
The rest of the $c_1$ terms are found in the complete NLO calculation 
\cite{BWHL}.

The off-diagonal one-loop elements of the soft anomalous dimension matrix 
are needed in the NNLO expansion at NNLL accuracy.
We find
\beq
\Gamma_{S\, 21}^{(1)}=\ln\left(\frac{u(u-m_t^2)}{s(s-m_t^2)}\right)\; ,
\quad \quad
\Gamma_{S\, 12}^{(1)}=\frac{C_F}{2N_c} \, \Gamma_{S\,21}^{(1)} \, .
\eeq

The two-loop soft anomalous dimension is calculated by analyzing all the 
relevant two-loop diagrams (c.f. \cite{NK2l,NKsch,NKtWH}). 
For the NNLO expansion at NNLL accuracy we need to determine the matrix element $\Gamma_{S\, 11}^{(2)}$. 
We find 
\beq
\Gamma_{S\, 11}^{(2)}=\frac{K}{2}\Gamma_{S\, 11}^{(1)}
+C_F C_A \frac{(1-\zeta_3)}{4}
\label{G112l}
\eeq
where $K$ is the two-loop constant defined previously. 
The two-loop result in Eq. (\ref{G112l}) is written in terms of the 
one-loop matrix element $\Gamma_{S\, 11}^{(1)}$, Eq. (\ref{Gamma111}). 

With these two loop results, we next calculate the NNLO soft-gluon corrections. 
The corrections, written in terms of the various coefficients and soft 
anomalous dimensions defined above, take the form 
\beqa
&& \hspace{-5mm}\frac{d^2{\hat\sigma}^{(2)}}
{dt \, du}
=F^B \frac{\alpha_s^2(\mu_R^2)}{\pi^2} 
\left\{\frac{1}{2} c_3^2 
\left[\frac{\ln^3(s_4/m_t^2)}{s_4}\right]_+ 
+\left[\frac{3}{2} c_3 \, c_2
-\frac{\beta_0}{4} c_3 +C_F \frac{\beta_0}{8}\right] 
\left[\frac{\ln^2(s_4/m_t^2)}{s_4}\right]_+ \right.
\nonumber \\ && \hspace{-5mm}
{}+\left[c_3 \, c_1+c_2^2-\zeta_2 \, c_3^2 
-\frac{\beta_0}{2} T_2+\frac{\beta_0}{4} c_3 
\ln\left(\frac{\mu_R^2}{m_t^2}\right)
+\frac{3}{2}C_F \, K-\frac{3}{16} C_F \, \beta_0 
+4 \Gamma_{S\, 12}^{(1)} \, \Gamma_{S\, 21}^{(1)} \right]
\left[\frac{\ln(s_4/m_t^2)}{s_4}\right]_+
\nonumber \\ && \hspace{-5mm} 
{}+\left[c_2 \, c_1-\zeta_2 \, c_3 \, c_2 + \zeta_3 \, c_3^2
+\frac{\beta_0}{4}  c_2 \, \ln\left(\frac{\mu_R^2}{s}\right)
-\frac{\beta_0}{2} C_F \ln^2\left(\frac{m_t^2-t}{m_t^2}\right)
-\frac{\beta_0}{2} C_F \ln^2\left(\frac{m_t^2-u}{m_t^2}\right) \right.
\nonumber \\ && \hspace{-5mm} \quad
{}-C_F \, K \, \ln\left(\frac{(m_t^2-u)(m_t^2-t)}{m_t^4}\right)
+B^{(2)}+3 D^{(2)}
+C_F\frac{\beta_0}{4} \ln^2\left(\frac{\mu_F^2}{s}\right)
-C_F\, K \ln\left(\frac{\mu_F^2}{s}\right)
\nonumber \\ && \hspace{-5mm} \quad 
{}+\frac{3\beta_0}{8} C_F \ln^2\left(\frac{m_t^2}{s}\right)
-C_F\left(\frac{K}{2}-\frac{3}{16}\beta_0\right) 
\ln\left(\frac{m_t^2}{s}\right)
+2 \Gamma_{S\, 11}^{(2)}
\nonumber \\ && \hspace{-5mm} \quad \left. \left. 
{}+\left(4 \, \Gamma_{S\, 12}^{(1)} \, \Gamma_{S\, 21}^{(1)} 
+4 \, (\Gamma_{S\, 11}^{(1)})^2\right) 
\ln\left(\frac{m_t^2}{s}\right)\right]
\left[\frac{1}{s_4}\right]_+  \right\} \, 
\label{NNLOts}
\eeqa
where $\beta_0=(11C_A-2n_f)/3$ is the lowest-order beta function.

This expression, Eq. (\ref{NNLOts}), extends the results in 
Ref. \cite{NKsingletopTev} from NLL 
to NNLL accuracy for $t$-channel single top production and it is 
the analog of the $s$-channel NNLL results in Ref. \cite{NKsch} 
and the $tW$ and $tH^-$ NNLL results in Ref. \cite{NKtWH}.

\mysection{Single top or single antitop $t$-channel production at the Tevatron}

We now use the previous theoretical expressions to study $t$-channel 
single top production at the Tevatron, noting that the results for 
single antitop production at the Tevatron are identical.
We add the NNLO corrections in Eq. (\ref{NNLOts}) to the NLO cross section 
and thus derive approximate NNLO cross sections at NNLL accuracy.
We use the MSTW2008 NNLO \cite{MSTW2008} parton distribution functions (pdf), 
as we also did in \cite{NKsch,NKtWH}.

\begin{figure}
\begin{center}
\includegraphics[width=11cm]{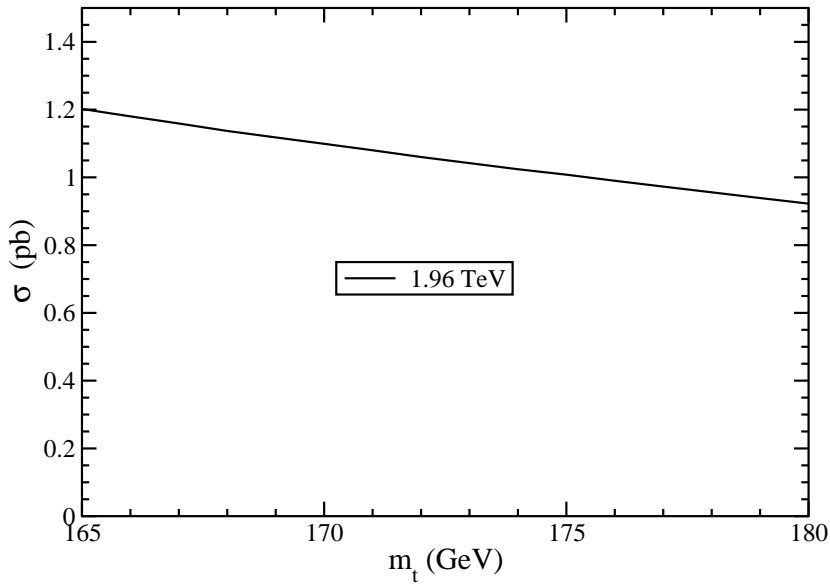}
\caption{The approximate NNLO cross section for single top quark production 
at the Tevatron with $\sqrt{S}=1.96$ TeV.}
\label{tevtop}
\end{center}
\end{figure}

In Figure \ref{tevtop} we plot the approximate NNLO cross section 
from NNLL resummation 
for $t$-channel single top production at the Tevatron with $\sqrt{S}=1.96$ TeV 
versus top quark mass in the range from 165 to 180 GeV.

\begin{table}[htb]
\begin{center}
\begin{tabular}{|c|c|c|c|} \hline
\multicolumn{4}{|c|}{NNLO approx single top $t$-channel cross section (pb)} \\ 
\hline
$m_t$ (GeV) & Tevatron 1.96 TeV & LHC 7 TeV & LHC 14 TeV \\ \hline
170 & 1.10 & 42.9 & 154 \\ \hline 
171 & 1.08 & 42.5 & 153  \\ \hline 
172 & 1.06 & 42.1 & 152 \\ \hline 
173 & 1.04 & 41.7 & 151 \\ \hline 
174 & 1.02 & 41.4 & 150 \\ \hline 
175 & 1.01 & 41.0 & 148 \\ \hline 
\end{tabular}
\caption[]{The single top quark $t$-channel production cross section in pb 
in $p \overline p$ collisions at the Tevatron with $\sqrt{S}=1.96$ TeV, 
and in $pp$ collisions at the LHC with $\sqrt{S}=7$ TeV  
and 14 TeV, with $\mu=m_t$ and using the MSTW2008 NNLO pdf \cite{MSTW2008}.
The approximate NNLO results are shown at NNLL accuracy.}
\label{table1}
\end{center}
\end{table}

Table 1 shows the numerical values of the cross section in pb for top quark 
mass values from 170 to 175 GeV in 1 GeV mass increments.
At Tevatron energy the NNLO soft-gluon corrections are positive and they 
increase  the NLO cross section by 4\%.
We note that the Tevatron cross sections presented in this paper are 
around 10\% smaller than those in \cite{NKsingletopTev}; that 
difference is mostly due to the new pdf used in this paper.

There are theoretical uncertainties associated with these values that 
arise from the dependence on the scale $\mu$ as well as from pdf errors.
The scale uncertainty is most commonly estimated by varying the scale 
by a factor of two, i.e. between $m_t/2$ and $2m_t$. For the approximate 
NNLO cross section at NNLL at the Tevatron the scale uncertainty is 
+0.1\% $-$1.8\%. 

The pdf uncertainty is calculated using the 40 different 
MSTW2008 NNLO eigensets as provided by MSTW at 90\% confidence level (C.L.)
\cite{MSTW2008}, which provides a conservative estimate of pdf error.  
For $t$-channel single top quark production at the Tevatron 
this 90\% C.L. pdf uncertainty is +6.0\% $-$5.9\%. 

The best current value of the top quark mass is $173$ GeV \cite{topmass}. 
For this top quark mass we write the $t$-channel single top quark cross section 
and its associated uncertainties expilicitly as 
\beq
\sigma_{\rm t-ch}^{\rm top}(m_t=173\, {\rm GeV}, \, \sqrt{S}=1.96\, {\rm TeV})
=1.04^{+0.00}_{-0.02} \pm 0.06  \; {\rm pb}
\eeq
where the first uncertainty is from scale variation and the second is 
the pdf uncertainty.

\mysection{Single top $t$-channel production at the LHC}

We continue with $t$-channel single top quark production at the LHC and we
present results at both 7 TeV and 14 TeV energies. We note that at the LHC 
the single top cross section is different from that for single antitop 
production. Here we study only single top, while in Section 5  
we study single antitop production.

\begin{figure}
\begin{center}
\includegraphics[width=11cm]{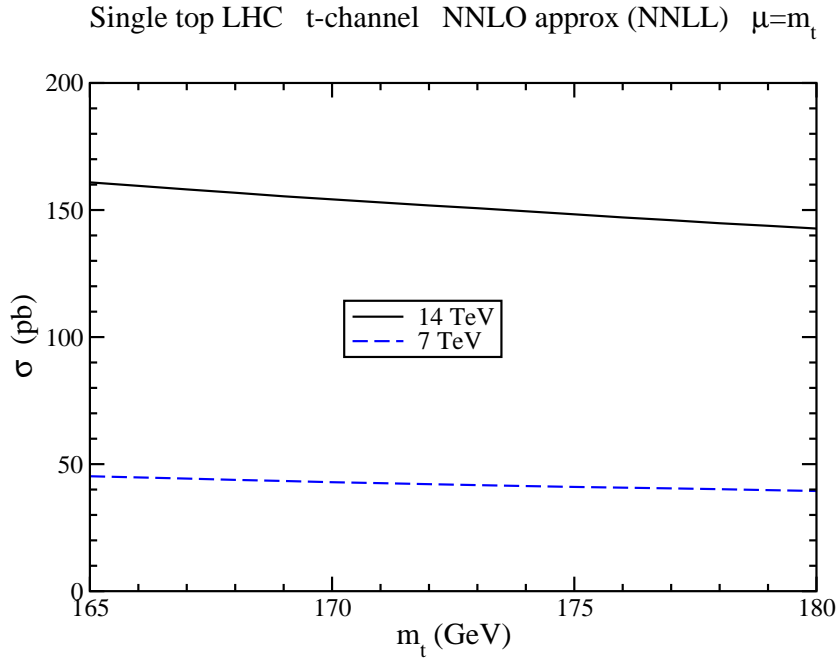}
\caption{The approximate NNLO cross section for single top quark production 
at the LHC with $\sqrt{S}=7$ TeV and 14 TeV.}
\label{lhctop}
\end{center}
\end{figure}

Figure \ref{lhctop} shows the approximate NNLO cross section 
from NNLL resummation 
for $t$-channel single top production at the LHC with $\sqrt{S}=7$ TeV 
and 14 TeV versus top quark mass in the range from 165 to 180 GeV.
Table 1 shows the numerical values of the cross section in pb for both energy 
values for top quark masses from 170 to 175 GeV.
At 7 TeV the NNLO soft-gluon corrections are negative and they decrease 
the NLO cross section by 1\%; at 14 TeV they decrease it by 3\%. 

At 7 TeV the scale uncertainty is $+3.8$\%  $-0.5$\% while 
the pdf uncertainty is $\pm 2.0$\% at 90\% C.L.
For $m_t=173$ GeV we have 
\beq
\sigma_{\rm t-ch}^{\rm top}(m_t=173\, {\rm GeV}, \, \sqrt{S}=7\, {\rm TeV})
=41.7^{+1.6}_{-0.2} \pm 0.8 \; {\rm pb}
\eeq
where the first uncertainty is from scale variation and the second 
from the pdf.

At 14 TeV the scale uncertainty is $+2.5$\%  $-0.6$\% while 
the pdf uncertainty is +1.8\% $-$2.2\% at 90\% C.L. 
For $m_t=173$ GeV we have 
\beq
\sigma_{\rm t-ch}^{\rm top}(m_t=173\, {\rm GeV}, \, \sqrt{S}=14\, {\rm TeV})
=151^{+4}_{-1} \pm 3 \; {\rm pb}
\eeq
where the first uncertainty is from scale variation and the second 
from the pdf.

\mysection{Single antitop $t$-channel production at the LHC}

We continue with $t$-channel single antitop quark production at the LHC and 
we present results at both 7 TeV and 14 TeV energies.

\begin{figure}
\begin{center}
\includegraphics[width=11cm]{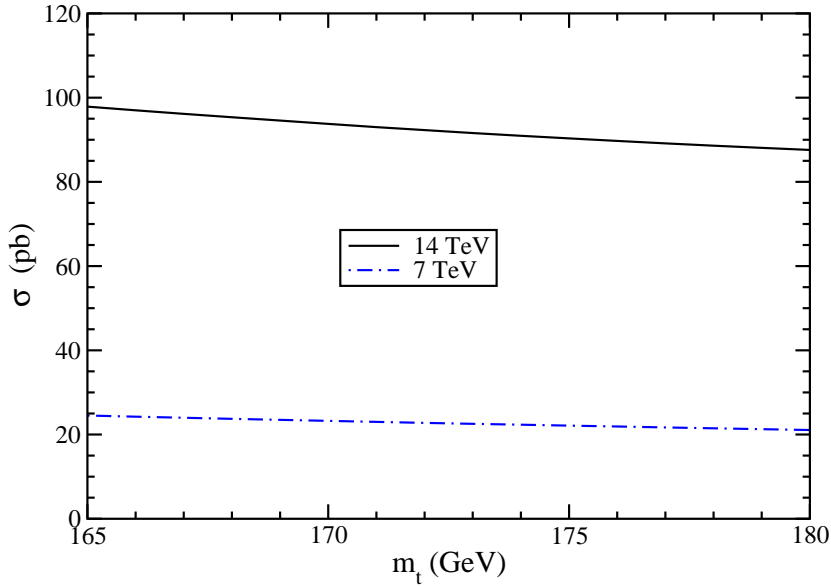}
\caption{The approximate NNLO cross section for single antitop production 
at the LHC with $\sqrt{S}=7$ TeV and 14 TeV.}
\label{lhcantitop}
\end{center}
\end{figure}

Figure \ref{lhcantitop} shows the approximate NNLO cross section 
from NNLL resummation 
for $t$-channel single antitop production at the LHC with $\sqrt{S}=7$ TeV 
and 14 TeV versus top quark mass in the range from 165 to 180 GeV.

\begin{table}[htb]
\begin{center}
\begin{tabular}{|c|c|c|} \hline
\multicolumn{3}{|c|}{NNLO approx single antitop $t$-channel cross section (pb)} \\ \hline
$m_t$ (GeV) &  LHC 7 TeV &  LHC 14 TeV \\ \hline
170 & 23.2 & 93.8 \\ \hline 
171 & 23.0 & 93.0 \\ \hline 
172 & 22.8 & 92.3 \\ \hline 
173 & 22.5 & 91.6 \\ \hline 
174 & 22.3 & 91.0 \\ \hline 
175 & 22.1 & 90.3 \\ \hline 
\end{tabular}
\caption[]{The single antitop $t$-channel production cross section in
$pp$ collisions at the LHC with $\sqrt{S}=7$ TeV and 14 TeV, 
with $\mu=m_t$ and using the MSTW2008 NNLO pdf \cite{MSTW2008}.
The approximate NNLO results are shown at NNLL accuracy.}
\label{table2}
\end{center}
\end{table}

Table 2 shows the numerical values of the antitop cross section in pb for both 
LHC energy values for top quark masses from 170 to 175 GeV in 1 GeV mass increments. At 7 TeV the NNLO soft-gluon corrections for single antitop production
are negative and they decrease 
the NLO antitop cross section by 1\%; at 14 TeV they decrease it by 3\%. 
This is the same percentage contribution as we found for single top production
in the previous section.

At 7 TeV the scale uncertainty is $+2.3$\% $-2.1$\% while 
the pdf uncertainty is $+3.0$\% $-4.0$\% at 90\% C.L.
For $m_t=173$ GeV we have 
\beq
\sigma_{\rm t-ch}^{\rm antitop}(m_t=173\, {\rm GeV}, \, \sqrt{S}=7\, {\rm TeV})
=22.5 \pm 0.5  {}^{+0.7}_{-0.9} \; {\rm pb}
\eeq
where the first uncertainty is from scale variation and the second 
from the pdf.

At 14 TeV the scale uncertainty is $+2.4$\%  $-1.0$\% while 
the pdf uncertainty is +1.9\% $-$3.2\% at 90\% C.L. 
For $m_t=173$ GeV we have 
\beq
\sigma_{\rm t-ch}^{\rm antitop}(m_t=173\, {\rm GeV}, \, \sqrt{S}=14\, {\rm TeV})
=92 {}^{+2}_{-1} {}^{+2}_{-3} \; {\rm pb}
\eeq
where the first uncertainty is from scale variation and the second 
from the pdf.

By comparing Figs. \ref{lhctop} and \ref{lhcantitop}, or Tables 1 and 2,
we see that the single antitop cross section in the $t$ channel is 54\% 
of that for single top at 7 TeV energy, while at 14 TeV it is 61\% of 
that for single top production. 

\mysection{Conclusions and combinations of results for all channels}

We have resummed collinear and soft gluon contributions to 
$t$-channel single top quark production at NNLL accuracy by using 
the two-loop soft anomalous dimension calculated in this paper.
We have expanded the resummed cross section to NNLO and provided 
numerical studies of the cross section at Tevatron and LHC energies. 
These NNLO corrections are small for the $t$ channel, in contrast to 
the much larger contribution found in the $s$ channel in Ref. \cite{NKsch}
and also for $tW$ (and $tH^-$) production found in Ref. \cite{NKtWH}. 

We can now present combined results for the $t$ and $s$ channels 
at Tevatron and LHC energies for a top quark mass of 173 GeV. 
At the LHC the $tW$ cross section is also sizable. 

For the Tevatron the sum of the cross sections in the $t$ and $s$ channels 
for single top production is $1.56^{+0.00}_{-0.02}  \pm 0.09$ pb, where the 
first uncertainty is from scale variation and the second is from the pdf. 
The cross section is the same for single antitop production at the Tevatron. 

For the LHC at 7 TeV the sum of the $t$ and $s$ channels for single 
top production is $44.9^{+1.6}_{-0.3} {}^{+1.0}_{-0.9}$ pb. For single antitop 
production the sum is $23.9 \pm 0.5 {}^{+0.7}_{-1.0}$ pb.
In addition the $tW^-$ cross section is $7.8 \pm 0.2 {}^{+0.5}_{-0.6}$ pb, 
and for ${\bar t} W^+$ it is the same as that for $tW^-$. 

For the LHC at 14 TeV the sum of the $t$ and $s$ channels for single 
top production is $159^{+4}_{-1} {}^{+3}_{-4}$ pb. For single antitop 
production the sum is $96 {}^{+2}_{-1} {}^{+2}_{-3}$ pb.
Also the $tW^-$ cross section is $41.8 \pm 1.0 {}^{+1.5}_{-2.4}$ pb, 
and the cross section for ${\bar t} W^+$ is the same. 

\mysection*{Acknowledgements}
This work was supported by the National Science Foundation under 
Grant No. PHY 0855421.

\end{document}